# Progress in Superconducting Metamaterials

Philipp Jung,[1,2] Alexey V. Ustinov,[1,3] Steven M. Anlage[2,4]

[1] *Karlsruhe Institute of Technology, Physikalisches Institut, 76131 Karlsruhe, Germany*

[2] *Center for Nanophysics and Advanced Materials, Physics Department, University of Maryland, College Park, MD 20742-4111 USA*

[3] *University of Science and Technology MISIS, Moscow 119049, Russia*

[4] *Electrical and Computer Engineering, University of Maryland, College Park, MD 20742 USA*

## Abstract

We review progress in the development and applications of superconducting metamaterials. The review is organized in terms of several distinct advantages and unique properties brought to the metamaterials field by superconductivity. These include the low-loss nature of the meta-atoms, their compact structure, their extraordinary degree of nonlinearity and tunability, magnetic flux quantization and the Josephson effect, quantum effects in which photons interact with quantized energy levels in the meta-atom, as well as strong diamagnetism.

## Introduction

The field of metamaterials entered its modern era at the dawn of the $21^{st}$ century and has continued to grow and develop in new ways. Metamaterials represent an organizing principle for developing artificial and engineered materials that create and enable unique interactions of matter with electromagnetic waves. Their realizations span the spectrum from DC to the ultra-violet. Their applications range from novel optics (e.g. negative refraction, hyper-lensing, cloaking, etc.) to sensing and detection, to quantum information science.[1,2,3]

As pointed out in our earlier review of superconducting metamaterials,[4] superconductors bring three unique advantages to the development of metamaterials, namely (A) low loss, (B) compact dimensions for the meta-atoms compared to other realizations, and (C) new opportunities for nonlinearity, tuning, and switching behavior. Since that review it has become clear that superconductors offer further unique advantages compared to ordinary metal, semiconductor and dielectric meta-atoms, namely (D) flux quantization and Josephson effects, (E) features arising from quantum interactions between photons and discrete energy states in the meta-atom, and (F) strong diamagnetism. The reader is urged to consult Ref.[4] for background discussion on superconducting physics relevant to metamaterials applications. In this article we shall review the progress in superconducting metamaterials made since the last review article was written (late 2010), with particular emphasis on features D-F mentioned above.

First we briefly review the key aspects of superconductivity relevant to this discussion. Superconductivity is a macroscopic quantum state of matter that arises from a many-body correlated electron state associated with Cooper pairing of electrons. Under many circumstances a superconductor can be described by a single coherent macroscopic quantum wavefunction with well-defined magnitude and phase. In the Ginzburg-Landau treatment, the superconductor is described in terms of a complex order parameter that has spatial and time dependence, and describes the response to variations in temperature, current, and magnetic field.[5] The magnitude of the superconducting order parameter is related to the density of superconducting electrons $n_s$, while its phase variation is related to the supercurrents. Superconductivity sets in as the equilibrium state below a transition temperature $T_c$. A superconductor can carry a finite amount of dissipation-free direct current, called the critical current $I_c(T)$, before it reverts to the normal state at temperature T. It can also exist in the presence of a magnetic field up to a lower critical field $H_{c1}(T)$ (before the onset of vortex entry) and an upper critical field $H_{c2}(T)$ before superconductivity is completely destroyed in the bulk (for type-II superconductors). Microscopically, an energy gap $\Delta(T)$ in the single-particle excitation spectrum opens up on the Fermi surface in the superconducting state. Magnetic fields are spontaneously excluded from the bulk of the superconductor through the Meissner effect. The fields are excluded from the superconductor through the development of screening currents that flow within a (microscopic) penetration depth $\lambda$ of the surface. The characteristic size of a Cooper pair is on the scale of the coherence length $\xi$, which also dictates the length scale on which the locally-varying superconducting order parameter heals due to a localized defect that suppresses superconductivity. Both the penetration depth and coherence lengths are microscopic, on the order of 10's to 100's of nm at temperatures well below $T_c$.

The complex conductivity $\sigma = \sigma_1 - i\sigma_2$, relating an electric field ($E$) to the resulting current ($J$) as $J = \sigma E$, satisfies the condition $\frac{\sigma_1}{\sigma_2} \ll 1$ for sufficiently low temperatures and frequencies. Hence superconductors have an electrodynamic response that is mainly inductive in character ($\sigma_2 > 0$) for sufficiently low frequencies and temperatures below $T_c$. The frequency limitation of superconducting response is set by the gap frequency $\sim 2\Delta(T)/h$, where $h$ is Planck's constant.[6] Beyond this frequency the electrodynamic properties approach those of the material in the normal state. Superconductors are lossy (i.e. they have non-zero $\sigma_1$) at finite frequency and temperature due to the existence of non-superconducting electrons (quasiparticles), whose abundance systematically increases with temperature.

This review is organized around the six unique properties that superconductors bring to the metamaterial field. These include (A) low losses due to the primarily inductive electrodynamic response of a superconductor, (B) the compact meta-atom structures enabled by the strong inductive response, (C) the strong nonlinearity and tunability brought about by superconductors at the extreme limits of their existence, (D) magnetic flux quantization and the Josephson effect, (E) quantum effects in which single photons interact with quantized states in meta-atom structures, and (F) the use of a strong diamagnetic response at low frequencies arising from the Meissner effect.

## A. Low-Loss Metamaterials

To be useful, metamaterials must have a low attenuation for the electromagnetic waves passing through them. Early work on superconducting split-ring resonators and related structures was reviewed previously.[4] Features such as evanescent wave amplification[7] and negative refraction are strongly suppressed with even small amounts of loss.[8,9,10,11,12] For example, the enhanced loss in metamaterials approaching the plasmonic limit has imposed a severe limitation on visible wavelength metamaterials composed of noble metal meta-atoms.[13,14,15,16] Recently it was pointed out that superconducting metamaterials show

the same plasmonic behavior, but as a function of temperature, rather than frequency.[17]

The low loss properties of superconductors are particularly clear in the THz frequency scale. In this case one must utilize high-temperature superconductors (HTS) that have a maximum superconducting gap frequency beyond 1 THz.[6] Split-ring resonators (SRRs) made up of HTS materials show a sharp onset of resonant interaction with THz radiation below the transition temperature.[18,19,20,21] Even low transition temperature superconducting SRRs are found to produce high quality factor resonances at frequencies of 0.14 THz,[22] up to 1 THz.[23]

## A1. Plasmonic Superconducting Metamaterials

The plasmonic behavior of superconducting meta-atoms has been examined by Kurter, *et al.*[17] They point out that below the transition temperature and gap frequency, a superconductor acts as a nearly loss-less plasmonic material with a temperature-tunable plasma frequency. As such, one can examine the plasmonic properties of superconducting metamaterials simply by varying temperature, and observe the approach to the plasmonic limit at essentially fixed frequency. They also calculate that the robustness of artificial magnetism in a magnetic meta-atom is greatly enhanced by superconducting electrodynamics. The condition for artificial magnetism in a superconducting meta-atom is given by the inequality $\frac{\sigma_1}{\varepsilon_0 \omega} < \frac{1}{2\pi} \frac{\lambda_{EM}/a_x}{1+R_p} \left(\frac{\lambda_{EM}}{2\pi\lambda(T)}\right)^2$, where $\sigma_1$ is the real part of the superconducting complex conductivity, $\omega$ is the angular frequency, $\lambda_{EM}$ is the electromagnetic wavelength, $a_x$ is the periodicity of the metamaterial, and $R_p = U_{kin}/U_{mag}$ is the plasmonic parameter which compares the energy stored in the kinetic energy of the carriers to the energy stored in magnetic fields of the resonant meta-atom.[16] This inequality is maintained over a broad temperature and frequency range, limited only by the critical temperature $T_c$ and gap frequency $\sim 2\Delta(T)/h$. The condition for artificial magnetism in a normal metal meta-atom is essentially identical except that the term $\left(\frac{\lambda_{EM}}{2\pi\lambda(T)}\right)^2$ is absent. We note that this term is the square of the ratio of a macroscopic length to a microscopic length, showing that artificial magnetism is significantly more robust in the superconducting meta-atom. In addition, superconducting metamaterials can typically be prepared with a much larger value of $\lambda_{EM}/a_x$ as compared to normal metal metamaterials (see section B below), making artificial magnetism even more robust.

Anagnostis, *et al.*, proposed a nano-scale superconducting plasmonic waveguide for THz radiation.[24]

Because of the Meissner effect, superconductors reject most of the electromagnetic fields from their interior, therefore they are not good as a single-surface plasmonic waveguide for tightly confining fields, except perhaps at temperatures very near $T_c$. However, putting a second superconductor nearby allows for a very compact field confinement. They calculate the effective index of refraction and attenuation length for a variety of plasmonic superconducting waveguide dimensions at THz frequencies.

Tsiatmas, *et al.* have studied the classic problem of radiation passing through an array of deep sub-wavelength-dimension holes.[25] It was found that superconducting electrodynamics alone explains most of the increase in extraordinary transmission of 75 GHz radiation through a sub-wavelength hole array in an HTS film. However there is an enhancement of transmission about 10 K below $T_c$, and this may be due to plasmonic interaction between the holes. Nearly simultaneous work by Tian, *et al.* examined extraordinary transmission at frequencies near 1 THz in HTS films with 50-65 μm scale hole arrays.[26] They found an enhancement of transmission through the hole arrays below $T_c$ and attributed it to resonant excitation of a real (as opposed to virtual) surface plasmon polariton mode in the array. It was proposed that the array could be used for low-loss large dynamic-range amplitude modulation and temperature controlled terahertz devices.[26]

### A2. Analogue Electromagnetically-Induced Transparency Metamaterials

A number of groups have recognized that meta-atoms can be combined in to meta-molecules in which the interactions between the atoms can give rise to qualitatively new phenomena. One of these phenomena is the creation of a classical analogue of the quantum effect of electromagnetically-induced transparency (EIT) in atoms using coupled classical resonators.[27,28,29,30]

The basic idea in classical EIT is to create coupling between two oscillators of dramatically different degrees of loss. The interference effect can create a strong transparency feature in the transmission spectrum, and can lead to a significant slowing of light in that frequency band. One approach is to employ asymmetrically split rings (ASRs) that have a 'dark' magnetic dipole mode that is coupled to a nearby 'bright' electric dipole mode, giving rise to Fano interference and a transparency window.[31] Superconductor / normal metal hybrid meta-molecules clearly offer the opportunity to create very strong classical EIT effects. The first work combined a lossy gold 'bright' resonator with a low-loss superconducting Nb 'dark' resonator.[32] A tunable transparency window with delay-bandwidth product on the order of unity was demonstrated.

The transparency window could even be switched off completely by increasing the intensity of the signal propagating through the meta-molecule.[33]

The EIT-like effects were extended into the THz domain by the Nanjing group. They have utilized NbN bright and dark resonators to interact with ps-duration pulsed radiation to create a transparency window having a delay-bandwidth product of 0.2.[34] By adding a second de-tuned dark resonator, they could create two transparency windows, although the performance is not as good as the single-resonator case.[34] Further work by the same group utilized an all-superconducting (NbN) coupled-resonator structure that showed enhanced slow-light features and superior transparency to the hybrid superconducting / normal metal (NbN/Au) structure of the same design.[35] The design allows simultaneous excitation of both resonators, and yields a delay-bandwidth product of 0.37. Recently, a hybrid HTS closed-ring resonator and Aluminum SRR system displayed a temperature tunable transparency window with a delay-bandwidth product on the order of 0.2.[36]

### B. Compact Meta-Atom Structure

Superconducting structures can be miniaturized to a large extent because the magnetic screening scale is microscopic (10's to 100's of nanometers), and because their $\sigma_2$-dominated electrodynamics means that losses do not scale adversely with reduced sample dimension, as discussed in detail previously.[4] This advantage has been exploited in a number of fields that utilize meta-atom-like structures for various purposes. Examples include optical detectors that exploit changes in kinetic inductance or resistance when a single photon is absorbed by a resonant superconducting structure, or superconducting qubits with transitions between quantized states in the microwave frequency regime (discussed in section E below).

Compact self-resonant superconducting spirals operating at radio frequencies maintain a high quality factor, $Q$, and function in the deep sub-wavelength regime where their physical dimensions are 1000 times smaller than the free space wavelength of the fundamental mode.[37,38,17,39] Laser scanning microscopy has been employed to image the current distributions in the spiral geometry,[38,40,41] and their electrodynamic properties are now well understood.[39] HTS spiral resonators with unloaded $Q$ values up to $10^3$ for the fundamental mode and 30 x $10^3$ for higher order modes have been demonstrated,[42] and their temperature and power tuning behavior explored.[43] These characteristics are comparable with natural atomic resonators such as the hydrogen atom, which has a similar size to wavelength ratio for visible light and possesses a $Q$ around 15 x $10^3$ for the Doppler broadened $H_\alpha$ line in the sun. Such high quality factors and deep subwavelength operation can only be attained through compact superconducting resonators. More compact superconducting resonators at even deeper sub-wavelength scales may be obtained by nano-patterning the spiral resonator in order to further enhance the kinetic inductance of the device and lower its resonance frequency. It should be noted that spiral meta-atoms were the platform for discovery of the anisotropic nonlinear Meissner effect in unconventional superconductors.[44]

Superconducting meta-atom-like structures have been developed by researchers in other areas for a number of different purposes. First, superconducting microresonators have been employed for microwave kinetic inductance detectors, and as bolometers, and often employ either a compact meandering co-planar waveguide design, or lumped inductors and capacitors.[45,46,47] The kinetic inductance optical detectors work by absorbing the photon which causes de-pairing in the superconductor, resulting in a change of the kinetic inductance and resonant frequency of the microresonator. The lumped-element designs allow for a larger collecting area, as compared to resonant transmission lines, because the current flow in the structure is more uniform and therefore the change in kinetic inductance due to absorption of an optical photon has a greater impact.[45]

Compact lumped-element superconducting resonators were developed earlier for use as backward-wave microwave filters.[48,49,50] Superconducting multi-mode composite SRRs have been used as microwave filters and show very low insertion loss and excellent out-of-band rejection.[51,52] More recently a novel compact fractal superconducting resonator has been developed utilizing only narrow superconducting lines, making it less susceptible to external magnetic fields.[53] This 6 GHz resonator was so small that it could be integrated on to a quartz tuning fork[54] and used for near-field scanning microwave microscopy.[55] Many compact self-resonant superconducting structures have been developed for use as quantum bits (qubits), as detailed in section E below.

## C. Nonlinearity, Tunability, Switching and Texturing of Metamaterials

Superconductors are fundamentally nonlinear electromagnetic entities. All superconductors are endowed with an intrinsic nonlinearity known as the nonlinear Meissner effect.[56,57,44] The Meissner effect is the spontaneous exclusion of magnetic flux from the bulk of an object when it undergoes the normal to superconducting phase transition, even in a static magnetic field. The material will achieve a lower free energy by entering the superconducting state (with a non-zero superconducting order parameter $\Psi$ and superfluid density $\sim |\Psi|^2$) and generating dissipationless screening currents to exclude the field, as long as the field is not too great. However, as the magnitude of the field is increased the superconductor must generate stronger screening currents (requiring a greater kinetic energy), resulting in a suppression of the superconducting order parameter, and the superfluid density. As a consequence, the superfluid plasma frequency[58] is decreased, and the magnetic screening length is enhanced. This in turn gives rise to an increase in inductance of any current-carrying circuit element made up of the superconductor, thereby generating a nonlinear electromagnetic response.[59,60] In general this intrinsic nonlinearity will be encountered when superconductors are pushed toward the limits of their existence in parameter space. This space is spanned by three quantities: temperature, applied magnetic field, and applied current. The limits of superconductivity are $T_c$ (the critical temperature), $B_c$ or $B_{c2}$ (the thermodynamic critical field or upper critical field for type-I and type-II superconductors, respectively), and $J_c$ (the critical current density). As superconductors are pushed towards these limits, the nonlinear properties will be strongly enhanced.

Superconducting metamaterials are particularly well suited for nonlinearity because they are typically made up of very compact (compared to the free space wavelength) meta-atoms (section B above), resulting in strong currents and fields within the meta-atom. The superconducting material at high current locations can then display nonlinear response because of the nonlinear Meissner effect, localized heating, or magnetic vortex entry. For example a current-tunable array of asymmetrically-split Nb resonators shows current-tunable transmission in the sub-THz band due to localized heating and vortex entry.[61] Alternatively, if nonlinear materials or devices are added at locations of extreme field or current, the meta-atoms can be very sensitively tuned by means of outside parameters.[62] The first example demonstrated was the change in superfluid density of a superconducting thin film induced by a change in temperature.[63] Later, it was demonstrated that the inductance of the superconductor changes significantly with the entry of magnetic vortices.[64] The change in resonant frequency of a Nb SRR with applied magnetic field was found to be quite complex and hysteretic.[64] Similar results showing complex tuning with magnetic field were later demonstrated at microwave frequencies using high-temperature superconducting SRRs[65] and at sub-THz frequencies with Nb SRRs of similar design.[66] The nonlinearity associated with the sharp resistive transition of the superconductor was exploited to bolometrically detect millimeter-wave radiation using the collective properties of arrays of Nb asymmetric split rings.[67] The temperature-tunable dielectric properties of superconductors play a central role in a proposed superconducting cloaking structure in the THz regime.[68]

The kinetic inductance of superconductors is strongly tunable, particularly in thin film structures (with at least one dimension on the order of the superconducting penetration depth) as a function of temperature. Such tuning has been accomplished at THz frequencies by varying temperature in HTS[20] and NbN thin film SRRs.[69] It was also demonstrated that the temperature tunability was enhanced by decreasing the thickness of the HTS films making up square SRRs.[70] THz time-domain experiments offer a unique opportunity to study fast nonlinear response in superconductors because the experiment involves a brief (~ ps) single-cycle THz excitation of the metamaterial. It was found that an intense THz pulse on a NbN metamaterial could produce significant de-pairing, resulting in a large quasi-particle density and increase in effective surface resistance of the film.[71,72,73] This in turn modulated the depth of the SRR resonance by 90%, without any significant change in temperature of the metamaterial. Such fast and large-dynamic-range tuning can be very useful for fast shutters or variable attenuators. The tuning of HTS SRRs with variable THz beam intensity could even be accomplished on ps time scales.[74]

### D. Magnetic Flux Quantization and Josephson Effect Metamaterials

Two other groups of metamaterials rely on an aspect of superconductivity that goes beyond the capability to carry an oscillating current with low loss. The macroscopic quantum wave function, or more specifically, its coherent phase, gives rise to two unique effects that open the door to a whole new class of capabilities and applications. It should be mentioned, however, that the quantum nature of these two effects, namely magnetic flux quantization and the Josephson effect, is purely macroscopic. For this reason, metamaterials employing the Josephson effect in conventional junctions and flux quantization in superconducting rings are described by classical dynamics, not quantum mechanics. True quantum metamaterials that mimic natural materials composed of meta-atoms with quantized energy levels can be assembled from arrays of qubits as described in section E below.

The flux-quantization condition states that the magnetic flux inside a superconducting loop can only be an integer multiple of the magnetic flux quantum $\Phi_0 = h/2e$. This is a direct consequence of the requirement that the phase of the macroscopic wavefunction has to change by an integer multiple of $2\pi$ on a closed loop path through the superconductor. To maintain this condition in the presence of an externally applied flux, the superconductor has to develop screening currents in the loop to generate a compensating flux to achieve the flux quantization condition. In a recent proposal[75] the authors suggest to match the line width and area of a superconducting loop in such a way that the currents necessary to screen one flux quantum exactly match the critical current of the loop. The resulting magnetization vs. applied flux curve of such an object would resemble a step function. By inserting it into the central area of a split-ring resonator, the authors aim to make the resonant response of this combined "flux exclusion" metamaterial nonlinear and quantized. Although they were able to fabricate an advanced "woodcut" type version of the meta-molecule and establish the basic principles of the concept, experimental constraints prevented them from reaching the parameter range necessary to demonstrate the desired effect.

The second effect related to the phase of the superconducting macroscopic wave function is the Josephson effect which occurs when two superconductors are separated by a Josephson junction, such as a thin tunneling barrier. In this case, the current through and voltage across the junction (I and V, respectively) are related to the phase difference between the two superconducting wave functions via the Josephson equations.[5]

$$I = I_{cJ} \cdot \sin(\delta)$$
$$V = \frac{\Phi_0}{2\pi} \dot{\delta}$$

Here, $I_{cJ}$ is the critical current of the junction, $\delta$ is the phase difference between the wave function on both sides of the junction, and $\dot{\delta}$ denotes its time derivative.

In the limit of small rf currents, the Josephson junction can be treated as a nonlinear and tunable inductor.[76] The value of the so called Josephson inductance

$$L_j = \frac{\Phi_0}{2\pi I_{cJ} \cos(\delta_0)}$$

depends on an additional dc current bias $I_0 = I_{cJ} \sin(\delta_0)$. One reason why this effect is so powerful in terms of tunability is that $L_j$ can be tuned from its initial value to infinity and even become negative.

Magnetic flux quantization and the Josephson effect come together when a superconducting loop is interrupted by one or more Josephson junctions. Here, the phase difference across the junction(s) depends on the externally applied magnetic flux. This is commonly known as a superconducting quantum-interference device (SQUID, or more specifically an rf-SQUID in the case of a single junction). Since the rf-SQUID meta-atom is a nonlinear and tunable analogue of the SRR,[77] we examine the electromagnetic properties of the SQUID near its self-resonant frequency. The original purpose of the rf-SQUID was to measure small magnetic fields and operate as a flux-to-frequency transducer.[78,79] The first proposal to use an array of rf-SQUIDs as a metamaterial was made by Du, Chen, and Li.[80,81] Their calculation assumes that the SQUID has quantized energy levels and considers the interaction of microwaves with the lowest states of the SQUID potential. For small detuning of the photon frequency above the transition from the ground state to the first excited state, the medium presents a negative effective permeability. The frequency region of negative permeability is diminished by a nonzero dephasing rate, and negative permeability will disappear for dephasing rates larger than a critical value.

rf-SQUIDs in the classical limit were modeled by Lazarides and Tsironis.[82] They considered a two-dimensional array of rf-SQUIDs in which the Josephson junction was treated as a parallel combination of resistance, capacitance, and Josephson inductance. Near resonance, a single rf-SQUID can have a large diamagnetic response. An rf-SQUID array displays a negative real part of effective permeability for a range of frequencies and magnetic fields. The permeability is oscillatory as a function of applied magnetic flux and is suppressed with applied fields that induce currents in excess of the critical current of the Josephson junction. Similar calculations, which included interactions between the rf-SQUIDs, were carried out by Maimistov and Gabitov.[83] They also consider the nonlinear wave equation for spatio-temporal waves in chains of SQUIDs. The response of a two-dimensional rf-SQUID metamaterial under a common oscillating magnetic field drive was also numerically investigated by Lazarides and Tsironis.[84] The weakly coupled elements of the metamaterial are predicted to show bistability and synchronization.

Recent experimental publications on single SQUIDs as meta-atoms[85,86] as well as on one-dimensional chains[87,88] and two-dimensional arrays[86] of SQUIDs have confirmed many of the theoretical predictions. It has been shown that in the case of small incident power levels, the SQUIDs in the array behave effectively as highly tunable and compact resonant meta-atoms. In addition to the tunability by temperature and magnetic field that is common in all superconducting resonators, their resonance can be very sensitively tuned using small DC magnetic fields at rates on the order of 10s of Terahertz per Gauss. Furthermore, this kind of tunability does not impact the quality of the resonance as severely as the other methods discussed in section C, because it does not rely on suppressing the superconducting order parameter.

One of the challenges when dealing with arrays made from large numbers of SQUIDs is keeping them uniformly biased. In a way, the benefit of the high degree of DC field tunability is also a curse, since even a small variation of the magnetic field across the sample can lead to a significant amount of detuning between the meta-atoms. Additionally, magnetic flux can get trapped inside the superconductor in the form of pinned magnetic vortices. This explains the requirement for rigorous magnetic shielding and proper sample design.[88]

Up to now, all the experiments on rf-SQUIDs and arrays have been performed in waveguides rather than in free space. Although this complicates the definition of material parameters, it has been shown that the effective relative permeability of a one-dimensional chain of SQUIDs coupled to a coplanar waveguide can drop below the critical value of -1.[87]

First measurements performed outside the quasi-linear regime at higher driving power levels[86,89] exhibit a rich spectrum of effects currently being investigated by several groups. The high degree of nonlinearity and the prospect of multi-stability[84,90] make this a strong candidate for potential SQUID meta-devices.

Apart from the notion of the SQUID serving as stand-alone meta-atom, a number of theoretical and experimental publications use two junction (dc-)SQUIDs to modify or enhance other metamaterial structures. These dc-SQUIDs have the advantage that they can be operated as field tunable inductors when inserted into a superconducting circuit as long as the currents flowing through them remain sufficiently small. In one case,[91] the authors incorporate a chain of several dc-SQUIDS into one of the loops of a double split ring resonator and demonstrated a resonance tunable with DC magnetic flux. In another paper,[92] the authors explore the idea of incorporating SQIFs (superconducting quantum interference filters) into superconducting spirals as a means to create a field tunable meta-atom that couples magnetically. Introducing Josephson junctions and/or SQUIDs into transmission lines also offers the possibility of engineering tunable band structures thus creating tunable, artificial one-dimensional crystals. This has been discussed theoretically[93,94] and shown experimentally[95] in a tunable composite left-right-handed transmission line.

A number of authors are using a similar concept to investigate other effects like parametric amplification,[96,97,98] impedance matching to a high-impedance source,[99] the dynamical Casimir effect (DCE)[100] or an analogue of Hawking radiation.[101] All of them are using chains of dc-SQUIDs in the central conductor of transmission lines or transmission line resonators and are based on the idea of modulating the electrical length (or speed of light) of the transmission line or resonator. In the case of the DCE, for example, the dc-SQUID array forms the center conductor of a superconducting co-planar waveguide ¼-wave resonator. The SQUIDs are biased with an AC flux at twice the resonant frequency while kept at low temperature (50 mK) such that quantum fluctuations dominate the electromagnetic properties of the resonator around its resonant frequency. Under these circumstances the pump converts virtual photons into real photons, which are measured and characterized by their frequency correlations. This nonlinear metamaterial thus facilitates the first conclusive experimental evidence of the DCE.

The use of a superconducting wire decorated with a series of Josephson junctions to create a tunable plasma edge metamaterial has not yet been explored experimentally. An array of such wires in a waveguide geometry, for example, will create a negative effective permittivity medium for frequencies below the wire-array plasma frequency.[102,103,104] Unlike a previous implementation with bare superconducting Nb wires,[105] the added Josephson inductance will allow for electrical current- or field-tunable changes in the plasma frequency over a substantial range. The recent development of 'superinductors,' made up of high linear density Josephson arrays, are well suited for this application.[106,107,107] Such structures are engineered to have a large tunable inductance with a minimum increase in loss.

Summarizing, Josephson devices remain one of the most intriguing and promising candidates for the future of superconducting metamaterials since they offer an unrivaled level of nonlinearity and tunability. Although experimental progress in this field is comparatively young, the increasing amount of research over the last few years is an encouraging indicator that this field is moving forward quickly.

## E. Quantum Metamaterials

One of the ways in which electromagnetic fields interact with natural materials is through coherent absorption and emission of photons by the atoms, which can be described as two-level quantum systems. Recently, artificially made quantum two-level systems have been reported in many experiments with superconducting nonlinear resonators cooled down to their ground state. At ultra-low temperatures, superconducting loops containing Josephson junctions behave as macroscopic two-level quantum systems often referred to as qubits.[108] The typical energy level separation of these superconducting quantum meta-atoms corresponds to a frequency which is on the order of a few GHz. The condition of keeping the thermal fluctuation energy $k_B T$ below the energy level separation $hf$ of the qubit with transition frequency $f$ in the GHz range requires temperatures well below 1 K.

In the past few years, research on superconducting qubits has made enormous progress in terms of design, fabrication and measurement techniques[108,109,110], which has led to orders of magnitude increase in coherence times and improved scalability. Although most of these impressive results have been achieved with the goal of quantum computing in mind, they also pave the way towards superconducting quantum metamaterials in which qubits serve as artificial meta-atoms.[111] In this case, the interaction between light and the metamaterial is described by photons coupling to artificial two-level systems (qubits). With typical transition frequencies in the microwave range, such meta-atoms can be seen as true scaled-up versions of natural atoms due to the quantum-mechanical nature of their interaction with the electromagnetic field.[109]

The notion of a quantum metamaterial suggests a treatment in a typical quantum-optical framework. In fact, many of the publications in this field over the last few years employ one of the most basic approaches in quantum-optics, namely an atom in a cavity. This system can be modeled using a Jaynes-Cummings Hamiltonian[112] which provides the standard description for an atom placed in a single-mode radiation field. Traditionally, this method is chosen in quantum electrodynamics (QED) and describes the enhancement of the coupling between atom and electromagnetic field by the resonance. Similarly, superconducting qubits can be placed in such a way as to strongly interact with photons in a microwave cavity. This technique is called circuit QED (cQED) and offers the possibility to read out the state of the qubit through the dispersive frequency shift of the resonator.

The theoretical frameworks for analyzing an ensemble of mutually non-interacting identical qubits, each of them coupled to a cavity, are the Dicke theory[113] and the Tavis-Cummings model.[114] The main result of the theory is the prediction of collective coupling of all N qubits to the cavity characterized by a $N^{1/2}$ enhancement of the individual qubit coupling strength. Such an enhancement has been observed experimentally for three superconducting qubits[115] and experiments on greater numbers of qubits are underway.

There are several theoretical publications discussing the physics of one-dimensional arrays of qubits coupled to transmission-line resonators,[116, 117,118,119] but so far there has been little experimental progress in the realization. The natural extension of this model is a finite chain[120,121] or network[122] of coupled cavities each of which in turn is coupled to an individual qubit (Jaynes-Cummings-Hubbard model). This case can be seen as a quantum photonic crystal and it bridges the gap to the multi-mode scenario of a transmission line or free space. Conversely, there has been a proposal for how to implement a metamaterial with specially tailored spectral density of modes into circuit QED.[123] By combining a section of left-handed transmission line with a right-handed transmission line resonator coupled to a qubit, ultra-strong multi-mode coupling can be achieved. By coupling qubits ("spins") to resonators ("bosonic modes") one can implement mapping to a variety of highly complicated and classically unsolvable Hamiltonians, which recently became an exploration approach by itself called quantum simulation. Superconducting quantum simulators[124] can be composed of arrays of microwave resonators and qubits coupled to them, which is a setting very similar to the quantum metamaterials discussed in this section.

Interaction of light with ensembles of two-level systems forming a quantum metamaterial generates non-classical photon states. This issue has been addressed in a number of theoretical papers dealing with superconducting quantum metamaterials[125,126]. There have been theoretical works for single qubits[127] and qubits spaced at intervals on the order of the wavelength in the gaps of a coplanar waveguide.[128,129] Experimental measurements of non-classical correlations for microwave photons are very challenging and became possible only recently[130,131,132], but no experiments have been reported for arrays of qubits up to this time.

In order to live up to the name of metamaterial, the research in this field has to break loose from its cradle of circuit QED. Proposed two-[133] and three-dimensional[134] superconducting quantum metamaterials made from networks of Josephson junctions may be the first steps in that direction. Also, more experimental implementations are needed to back up the many theoretical works and proposals. Several experiments[135,136,137] have already demonstrated the possibility of measuring individual superconducting qubits in a transmission line, without using a resonator. We will undoubtedly see experiments with arrays of superconducting qubits placed in transmission lines or waveguides in the near future.

Superconducting quantum metamaterials containing large numbers of artificial atoms offer a wide range of prospects, from detecting single microwave photons[138] to quantum birefringence[134] and superradiant phase transitions.[139] In contrast to natural atoms or molecules, superconducting qubits allow for a very strong effective dipole coupling to the external electromagnetic field. This opens up unique opportunities of designing artificial quantum structures made of meta-atoms that have ultra-strong and coherent coupling to the electromagnetic fields in a transmission line or a cavity. The major technical challenge for artificially made quantum metamaterials will be making the qubits as identical as possible as the energy level separation varies for physically different meta-atoms. This problem can be circumvented by utilizing very strong coupling of the meta-atoms to the electromagnetic field, similar to the way of overcoming the effects of inhomogeneous broadening in lasers made of natural atoms. This should allow for novel ways of generating and controlling non-classical electromagnetic waves (light squeezing, coherent down- and up-conversion, etc.). This emerging field will be driven by many interesting new experiments in the near future.

## F. Diamagnetic Metamaterials

Inspired by the proposal of a DC magnetic cloak[140] and early experiments on implementations using superconductors,[141] there has been a lot of progress in the field of magneto-static transformation optics. While traditional shielding techniques using materials of either very high or low permeability (i.e. μ-metal and superconductors) can shield an object from an external field, they also distort the field lines thus making the object detectable. This can be avoided by using a shell made from a magnetically anisotropic material. The original concept, however, required the magnetic permeability to take extreme values near the inner and outer edges of the shell. In a more recent proposal,[142,143] however, it has been shown that good cloaking can be achieved using a cylindrical shell with a locally uniform, anisotropic $\mu$. The condition required to achieve this effect is only that the radial and angular component of the relative, magnetic permeability, $\mu_\rho$ and $\mu_\Theta$, have to fulfill the equation $\mu_\rho * \mu_\Theta = 1$. This ensures that the field outside the shell remains essentially unaltered by the presence of the shielded object. The fields inside the shell, however, also depend on the ratio between $\mu_\rho$ and $\mu_\Theta$: If $\mu_\rho \gg \mu_\Theta$, the fields are concentrated on the interior of the shell. There is a proposal to implement this using superconducting and ferromagnetic wedges or prisms for magnetic field concentration.[144,143] In the opposite case, $\mu_\Theta \gg \mu_\rho$, the interior of the shell remains essentially field-free. According to,[142] this classic cloaking case can be realized similar to the original proposal[140] by combining paramagnetic layers with superconducting plate arrays or using arrays of superconducting and ferromagnetic strips.[145] An additional superconducting layer on the inside of the shield could be used to ensure that fields from sources inside the shell cannot leak out. A design based on the same concept has also been proposed as a carpet cloak.[146]

The first experimental implementation of a cylindrical DC magnetic cloak[147] was made using superconducting Pb plate arrays and permalloy layers similar to the original proposal by Wood and Pendry.[140] Local field measurements revealed a good shielding of the externally applied magnetic fields on the inside while the fields outside the shield, close to its surface were almost completely unaltered by the presence of the cloak over a range of magnetic field values.

A much simpler design has been implemented[148] using only two coaxial cylinders the inner and outer made from rolled up superconductor and ferromagnetic foil, respectively. This design is only suitable for uniform external magnetic fields, but doesn't require any form of micro fabrication and can be built from commercially available products. In the presented field probe measurements, the authors demonstrated a significant, albeit not complete, cloaking effect. Further studies on this design[149] revealed that it can maintain some of its cloaking properties for very low frequencies. This AC effect, however, is significantly disturbed by the hysteretic behavior of the ferromagnet.

## Discussion and Conclusions

In Fig. 1 we present a summary of realized and notional superconducting meta-atom and meta-molecule styles. The meta-atoms are grouped in terms of their primary response. Fig. 1(a) shows mainly magnetically active meta-atoms, although the split-rings can also be excited electrically. The realizations include (from top-to-bottom) single and double[105] split rings, two types of asymmetric split rings,[31,22] the split-cross structure,[18] the woodcut flux-quantization split-ring resonator,[75] the spiral,[37] and rf-,[80],[85] dc-,[97] and multi-junction[136] SQUIDs and qubits. Fig. 1(b) shows electrically active meta-atoms ranging from the bare wire[105] to Josephson-decorated wires, to wires augmented with field-[91] or current-tunable dc-SQUIDs. Fig. 1(c) shows implementations of the Josephson transmission line and the conventional composite right-/left-handed transmission line[49] and its tunable version utilizing dc-SQUIDs.[95] Fig. 1(d) shows superconducting lumped-elements acting as meta-atoms.[46,48] Fig. 1(e) shows several types of meta-molecules, including one creating a classical analog of electromagnetically-induced transparency,[32] a superconducting quantum interference filter structure,[150] and a dc-SQUID array transmission line.[96] Fig. 1(f) shows two realizations of dc magnetic cloak structures.[141,147]

The future for superconducting metamaterials looks particularly bright. In the low-loss category, one can expect new plasmonic structures that can guide tightly-confined electromagnetic waves over significantly longer distances than normal metal structures. The addition of Josephson tunability combined with the low losses of superinductors and bulk superconductors should allow the realization of strongly tunable devices that retain their low loss properties, a key goal for quantum information processing and electromagnetically-induced transparency with high delay-bandwidth product. We also anticipate new results displaying bistability and multi-stability in rf-SQUID arrays, testing the theoretical predictions. In the compact structure category we can expect to see even smaller meta-atom designs as Josephson and SQUID inductances are included in various split-ring and spiral resonator designs, as well as transmission line realizations. The initial round of experiments on superconducting quantum metamaterials should come to fruition in the near future. Quantum metamaterials comprised of arrays of superconducting qubits are an emerging new field for fundamental studies in quantum optics, opening the possibility to explore collective quantum dynamics under very strong coupling between electromagnetic fields and artificial atoms. We also anticipate a convergence of metamaterial ideas and quantum simulations as the number of qubits is scaled up and concepts of modeling various complex Hamiltonians are further explored.

**Acknowledgements**

This work is supported by the NSF-GOALI and OISE Programs through Grant No. ECCS-1158644, the Center for Nanophysics and Advanced Materials (CNAM), the EU project SOLID, the Deutsche Forschungsgemeinschaft (DFG), and the State of Baden-Wurttemberg through the DFG-Center for Functional Nanostructures (CFN) within subproject B3.5. We thank Susanne Butz for helpful discussions. P.J. acknowledges the support by the Karlsruhe House of Young Scientists (KHYS) through the KHYS research travel scholarship.

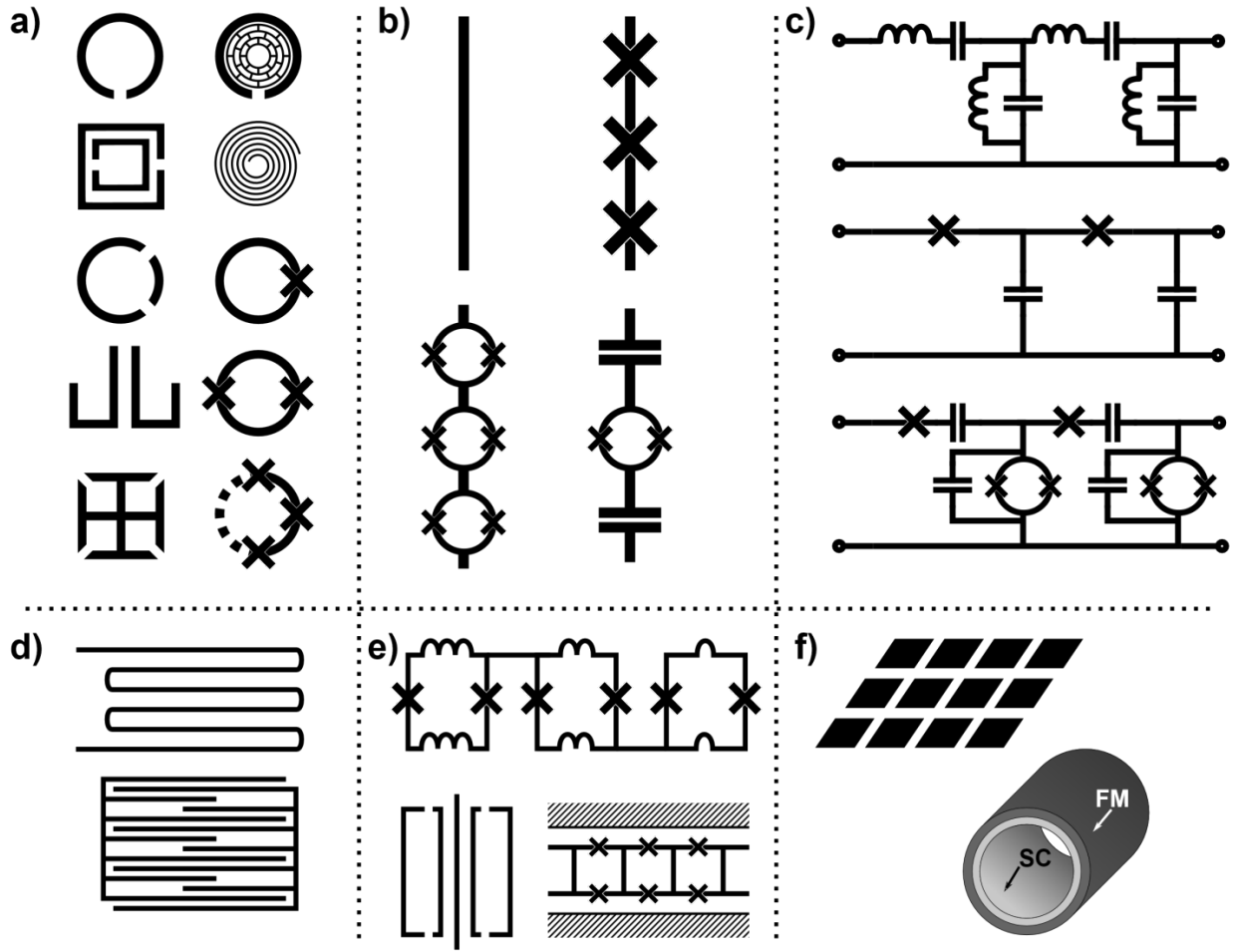

Fig. 1 Semi-schematic representation of a subset of superconducting meta-atoms and meta-molecules. a) Magnetically active meta-atoms, b) electrically active meta-atoms, c) transmission line metamaterials, d) lumped-element meta-atoms, e) various meta-molecule realizations, and f) dc magnetic cloak structures.